\documentclass[letterpaper,twocolumn,showpacs]{revtex4}
\usepackage{graphicx,bm}
\begin{document}

\title{X-ray tomography of a crumpled plastoelastic thin sheet}

\author{Yen-Chih Lin$^1$, Ji-Ming Sun$^1$, H. W. Yang$^1$, Yeukuang Hwu$^2$, C. L. Wang$^2$, and Tzay-Ming Hong$^1$}
\affiliation{$^1$Department of Physics, National Tsing-Hua University, Hsinchu 300, Taiwan, Republic of China\\
$^2$Institute of Physics, Academia Sinica, Taipei 115, Taiwan, Republic of China}
\date{\today}



\begin{abstract}
A three-dimensional X-ray tomography is performed to investigate the internal structure and its evolution of a crumpled aluminum foil. The upper and lower bounds of the internal geometric fractal dimension are determined, which increase with the compression. Contrary to the simulation results, we find that the mass distribution changes from being inhomogeneous to uniform. Corroborated with the evidence from previous experiments, these findings support the physical picture that the elastic property precedes the plastic one at dominating the deformation and mechanical response for all crumpled structures. 
We show that the interior of a crumpled ball at the plastic regime can be mapped to the compact packing of a granular system.
\end{abstract}

\maketitle

\section{Introduction}

Crumpling is a common but complicated process that occurs in many occasions and at all length scales. It is more complex in sheets than wires\cite{Herrmann} because the former develops a unique cone structure. Although much has been learned from studying the mechanism of a single cone\cite{Witten07, Mahadevan99}, the collective behavior of vertices is still an open field. The main difficulty lies in the treatment of topological constraint that they cannot cross each other. The  first aim of this Letter is to acquire 3D images of this exotic self-organized crumpled structure. The conventional process to achieve this purpose is to unfold and profile the topography of its surface, via which the probability of different ridge length was found to obey the lognormal distribution at large scale and a power law at small scale\cite{Kudrolli}. However, so far there is no systematic study on the evolution and statistical properties of its labyrinthical interior without destroying the crumpled ball. 


Another interesting property is the scaling relation proposed by Nelson and Kantor\cite{Nelson_book}, who predicted the value of dimension $D$ appearing  in $R\propto R_{0}^{2/D}$ where $R_0$ and $R$ are the initial size and final radius of the crumpled ball. The development of this theory has relied mainly on computer simulations over the years. First, Vliegenthart and Gompper\cite{Gompper} studied the elastic sheet and extended the scaling relation to include the external force $F$ and various modulii:
\begin{equation}
\label{scaling}
\frac{R}{R_{0}}\propto\left(\frac{K_{0}R_{0}^{2}}{\kappa}\right)^{\beta}\bigl{(}\frac{\kappa}{FR_{0}}\bigl{)}^{\alpha}
\end{equation}
where $K_{0}$ and $\kappa$ are the two-dimensional Young's modulus and bending rigidity, and the exponents  $\alpha$ and $\beta$ are believed to be universal and material-independent. More recently, Tallinen {\it et al.}\cite{Timonen} included plasticity but still found this scaling relation to remain intact. 
Furthermore, ${\AA}$str${\ddot{o}}$m {\it et al.}\cite{jam} reported  a divergence in $F$ when the volume fraction of the sample is around 0.75, at which point the elastic behavior turns out to be similar to that of a granular media. 
In the last decade, scientists have performed many experiments to determine the dimension $D$ of elastic and plastoelastic crumpled structures\cite{Balankin1, Balankin2, Gomes1, Gomes3}. Most were based on the scaling relation in Eq.(\ref{scaling}) to give this dimension as a function of the two universal exponents, $D=2/(1+2\beta -\alpha )$\cite{Balankin1, Balankin2}. The fractal dimension is believed to be material dependent but, otherwise, is universal for the external force and different initial size, thickness, and final radius of the sample.

It is obvious that a direct observation of the internal structure requires a nondestructive method that can see through the crumpled ball with sufficient resolution to describe the exact location of thin sheets. Normal destructive approach, such as cutting the ball in half to expose the cross section is not only affected by possible alteration of the detailed structure, but also limited severely by the number of planes one can expose and therefore reduces the possibility to reconstruct 3D structures.  X-ray microtomography is certainly the experimental tool of choice to meet these two demanding requirements simultaneously.

\section{Experiments}
 
We use the aluminum foils as our sample, which are of the same thickness (16$\mu$m) but different radius $R_{0}$ [mm]=3, 4, 4.5, 5, 6, 6.5, 7, 8, 9, and 10. They are randomly folded into a ball of same final radius $R=$1.5mm.  Such a dimension is already difficult to be examined by other methods, and hard-X-ray with photon energy higher than ~10keV is required.  To precisely describe the crumpled geometry of the foil with a spacing between adjacent sheets in the order of $\mu$m, high image resolution and precision of specimen placement are essential. 

In order not to sacrifice too much resolution, the sample is limited to $5\times 5\times 5$ mm$^3$ for the x-ray tomography. This makes the preparation of sample by pressure chamber\cite{Neil1} unrealistic. After trying several tries, we eventually settled by mimicking the method first employed by Balankin {\it et al.}\cite{Balankin1}; namely, use a flat tip conventional tweezer to sqeeze while rotating the sample. The method is believed to produce reliable and reproducible data.

Under the limitation of beam time, we judge that it is more preferable to invest it on samples with differerent parameters. Many anomalous properties have been revealed in the literature, which were sensitive to the compaction $R/R_{0}$ of the sample. For instance, there was a two-stage transition (folding-crumpling transition) in the intermediate range of compaction around 0.23\cite{Timonen2}. Furthermore, a jamming transition was found when the sample reached the highly compact state. Therefore, a detailed scan of different compactions can be informative.

 We employ a special version of X-ray microtomography system based on the high intensity X-ray from synchrotron\cite{hwu1}. Such systems provide a standard resolution between 1$\mu$m to 2$\mu$m which shows clear reconstructed images for our analysis. The experiment is performed at the 01A beamline of National Synchrotron Radiation Research Center (Hsinchu, Taiwan).  The beamline provides unmonochromatic X-rays  whose energy distribution is 8keV to 15keV. Image acquisition time per projection is about 10ms, which is captured by a CCD with 2X optical lens focused on a CdWO$_{4}$ single crystal scintillator. The resulted reconstruction consists of a data matrix of 1200$\times$1200$\times$1200 pixels with a pixel size 3$\mu$m.  Higher resolution image can be obtained by transmission X-ray microscopy which has recently achieved a resolution of 30nm\cite{hwu2}. However, the high resolution comes with a trade-off with reducing specimen size.

\begin{figure}
\begin{center}
\includegraphics[width=8.5cm]{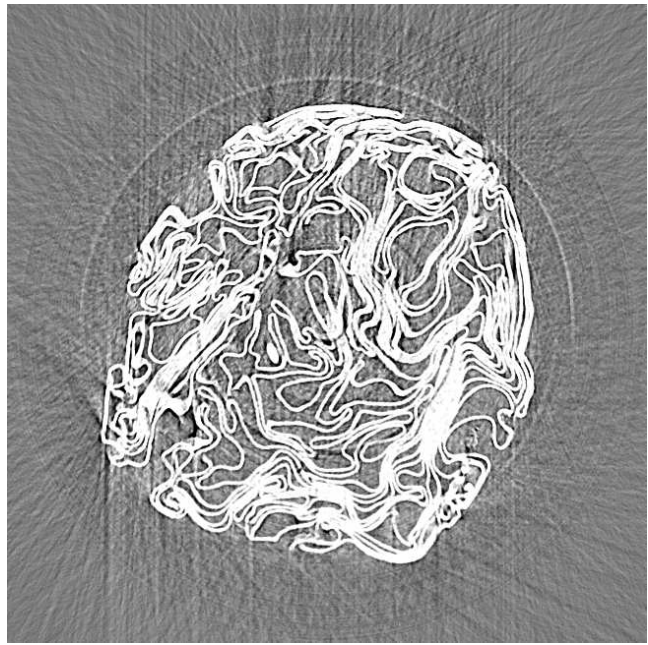}
\caption{A slice of raw images reconstructed from 1000 projections for $R/R_0 =0.167$.  Before segmentation, two kinds of reconstructed artifacts are present. The glisterns are caused by the strong refraction and diffraction of incident X-ray from aluminum facets when they are aligned to the incident beam within a certain value. The ring artifacts are due to the defects on the detector screen and insufficient background normalization.  Both kinds of noise can be reduced significantly by the fill tracing method. } \label{fig:TomoRC0589}
\end{center}
\end{figure}

\begin{figure*}
\includegraphics[width=1\textwidth]{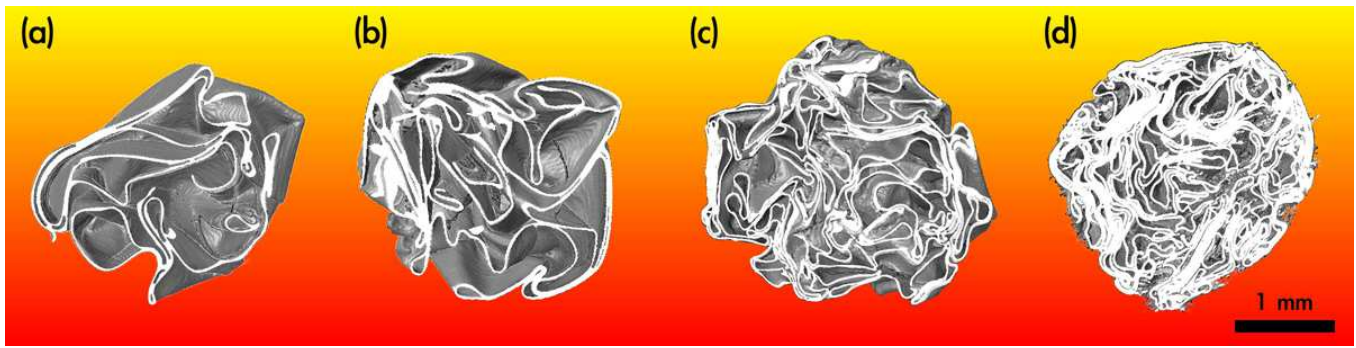}
\caption{(color online) Sequence photos of the reconstructed cross section of crumpled samples. The compactions $R/R_{0}$ of (a), (b), (c) and (d) are 0.38, 0.30, 0.19 and 0.15, respectively. As the compaction decreases, more vertices are induced  and more locally-aligned structure appears. }\label{fig:recon}
\label{recon}
\end{figure*}

\section{Data reconstruction}

The tomography reconstruction is done with interpolation on pixels which is not covered during the rotation. It is therefore important to acquire enough number of projections experimentally to perform the reconstruction. However, the experimental system, the aluminum foil here, is quite simple and its thickness can be assumed not to change during the compaction, then the 
interpolation is not likely to create problem during the filtered-back projection methods for reconstruction.  There are quite a few more algorithms scientists use to get the tomography reconstruction, but we believe that they do not affect the result in this case.  The only possible problem is that, due to the resolution limit, if the foil wrinkled with scale smaller than the 
resolution then the reconstruction will not be able to reproduce it and that could affect the numerical analysis. This has been checked by opening the foil and confirming that there was no 
small scale distortion. 

In image segmentation, the conventional thresholding is an essential and widely-used method to spot the object pixel and convert gray-scale images into binary ones. Despite its simplicity, the only tunable parameter, the threshold value, constrains the precision of the method. Inevitably, some useful information is erased during this thresholding of the raw data. To minimize the loss, the original images are converted into 1-bit units through the fill tracing process\cite{seg_book}, which makes use of vectorizing to trace out the interface between the object and the background. To be rid of the tiny areas due to noise and dusts, see Fig.\ref{fig:TomoRC0589}, minimum area is set at 20 pixels to filter out the invalid tracing. Although this method can not distinguish two paths when they are in contact, it provides a sufficient quality to the information for our analysis. All data matrices are reconstructed again after the preliminary analysis has been processed. Sample pictures are shown in Fig.\ref{fig:recon}.

The reason why the thresholding is required is that the imaging process in the experiment does not give the absolute gray scale. Every projection is composed of dark noises (from CCD, electronic noise, etc.), background noises (from the optical elements such as the X-ray windows, scintillators, lens inferfections, etc.) and other effects. Therefore, it is hard to get the absolute density.  For example, in principle we should be able to extract the absorption coefficient of aluminum from the image gray scale comparing to air, but that is normally not reliable unless a large calibration is done. In this case, we do not care about the absolute value and our case becomes practically a 1-bit system. The only reason to perform thresholding is to eliminate noise and artifacts from the reconstruction and more subtly the phase contrast effect, such as those lines tangential to the edge of the object. The danger of thresholding is the elimination of small details like those wrinkles. One can answer this concern by blowing up the reconstructed image and show the effect of different thresholding and its effect on image.

\section{Internal geometric fractal Dimension}

Fractal dimension is a useful quantity to characterize many cluster-assembled structures and the surface of corrugated thin films with self-affine morphology\cite{Cannell}. However, cares need to be taken when comparing the internal geometric fractal dimension $D_m$ with the dimension $D$ determined from its mass-size or scaling relation in Eq.(\ref{scaling}). For all classic fractals, such that Koch curves, Sierpinski gasket and sponge, etc., the 
equality $D_m = D$ is obvious from construction. 
In the case of forced folding, the fractal dimension of the set of 
balls folded from sheets of different sizes is determined by the 
folding conditions. For example, the set of balls with the same 
contraction ratio $R_0/R$ has the fractal dimension $D = 2$. 
At the same time, the sets of balls folded by the same force ($F$= 
constant) and under the same stress pressure ($P \propto F/R^2$ 
=constant) are characterized by different fractal dimensions $D_P > 
D_F > 2$, whereas the internal structure of each folded ball does not know 
how the other balls were folded.
Furthermore, in the case of (hyper) elastic sheets as, for instance, 
rubber sheets, the balls are completely unfolded after the confinement 
force is withdrawn. In contrast to this, the diameter of folded paper 
sheet increases only slightly increases after withdrawing. So, the 
fractal dimension of the set of folded paper balls is dependent on the 
strain relaxation rate. 

Balankin {\it et al.}\cite{Balankin2} have employed the mthod proposed by Miyazima and Stanley\cite{stanley} to measure $D_m$ by piercing a needle connected to a string along the diameter of the crumpled ball and studying the number of intersections as a function of the sheet size. They found that the external structures of balls folded from 
different papers are characterized by the same fractal dimension $D$, 
i.e., $D$ is independent on the mechanical and geometrical properties 
of paper sheet. Furthermore, they noted that 
this dimension was numerically close to the value of $D$ 
expected from numerical simulations for the set of phantom sheets 
folded under a fixed force instead of a fixed pressure. 
From this observation they speculated that the empirical 
relation between force and ball diameter in Eq.(\ref{scaling}) was valid not only for sheets 
of different size $R_0$, but also for force distribution within the 
folded ball.
In their experiments, the ratio $R_0/R$ is relative small in the sense that the self-avoiding effects do not play a significant role. When the ball is further compressed, the difference between the self-avoiding paper and the phantom sheets may become more pronouced. 

\begin{figure}
\begin{center}
\includegraphics[width=8.5cm]{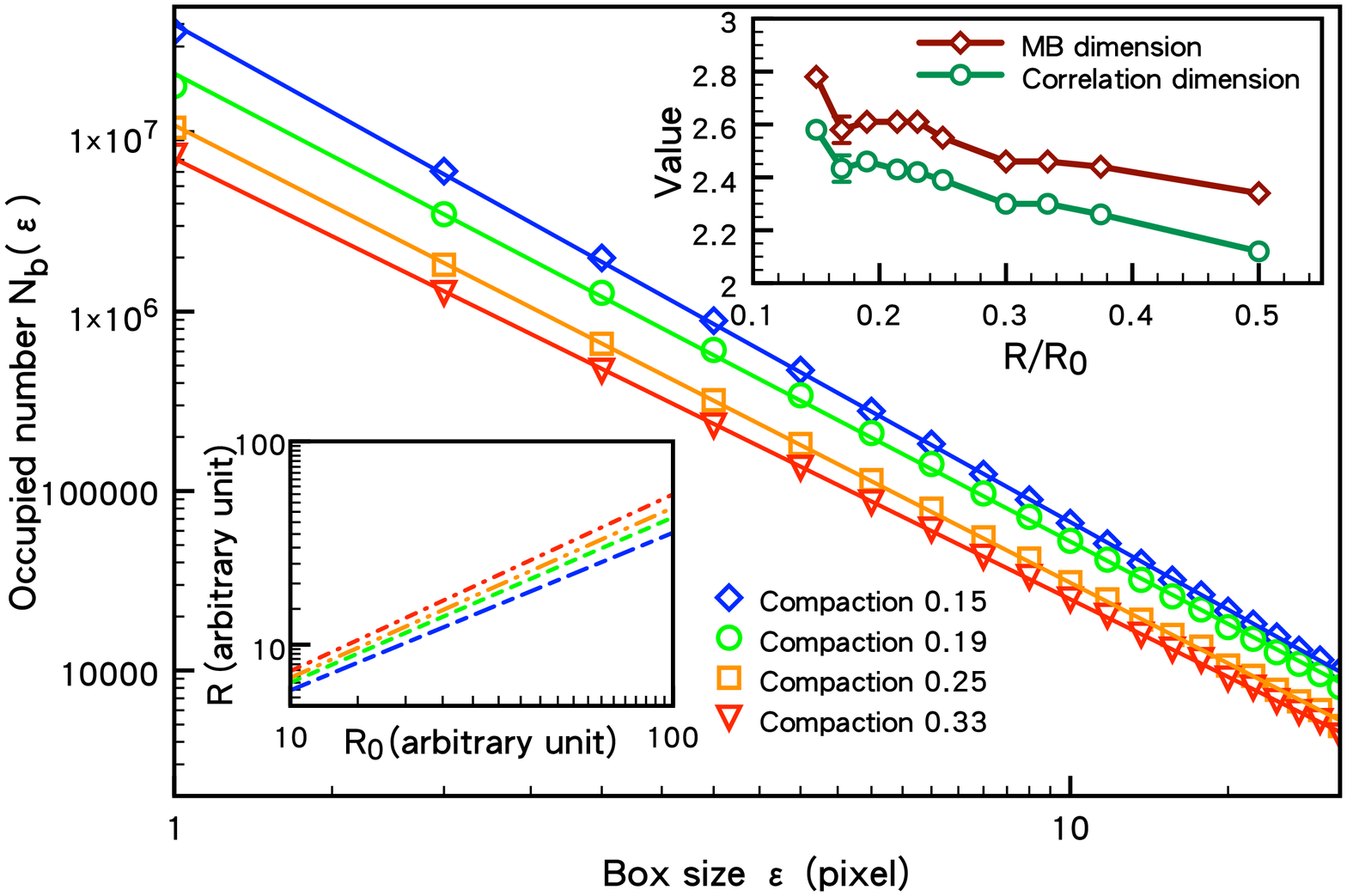}
\caption{(color online) The occupied number of boxes is plotted as a function of box size in a log-log scale. Its slope can be readily read off as the MB dimension. Together with the correlation dimension, they are shown in the right inset for different compaction $R/R_0$. These two dimensions set the upper and lower bounds for the fractal dimension. We estimate it to increase roughly from 2.2 to 2.8 as larger sheets are used to achieve lower compactions. The left inset is a log-log plot of $R$ versus $R_{0}$ according to the scaling relation of Eq.(\ref{scaling}). The slope of different lines is adopted from the experimental values in the right inset. To examine the reproducibility of the measurement, a representative error bar in the inset is drawn for $R/R_{0}$=0.17 from three more samples. The crumpled ball at this ratio just crossed the loose packing regime into the compact one. The smallness of error justifies the reliability of our data.} \label{fig:fractal}
\end{center}
\end{figure}

Here, we present a more direct measurement of $D_m$ by adopting the box-counting method to calculate the Minkowski-Bouligand (MB) dimension, $D_{b}\equiv -\lim_{\epsilon_{b}\rightarrow 0}(\log N_{b}(\epsilon_{b}))/(\log\epsilon_{b})$. In Fig.\ref{fig:fractal}, the number of occupied boxes, $N_{b}$, decays with the box size, $\epsilon_{b}$, in a power-law fashion, and the slopes of these solid lines are the  MB dimension. Since the value of $D_{b}$ is often considered to be the upper bound for the fractal dimension, we also calculate the density-density correlation to determine the correlation dimension to set the lower bound. The correlation dimension is defined as $D_{c}\equiv -\lim_{\epsilon_{c}\rightarrow 0}(\log C(\epsilon_{c})/(\log\epsilon_{c})$ where the correlation function $C(\epsilon_{c})$ counts the number of pairs $(i,j)$ with length $s(i,j)<\epsilon_{c}$. 
From the right inset in Fig.\ref{fig:fractal}, we find that both MB and correlation dimensions grow as the compaction $R/R_0$ decreases. 

\section{Mass distribution}

In the crumpling of an elastic wire, the mass-size relation has been studied and compared to the Flory's theory for polymers\cite{Neil2, cw, Gennes}. Adopting Flory's free energy expression, it was found that the bending and exclusion energies dominate the system rather than the entropy\cite{Neil2}. This is equivalent to saying that the crumpling is a deterministic process and all the uncertainties of the final configuration are attributed to the fluctuation and imperfection during the compression.

\begin{figure}
\begin{center}
\includegraphics[width=8.5cm]{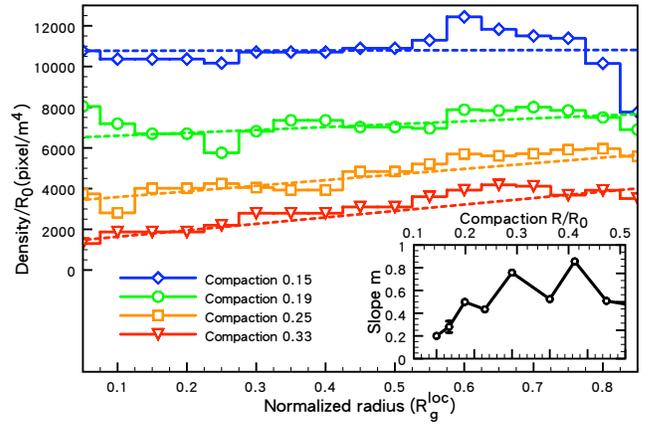}
\caption{(color online) The mass density distribution for four different compactions $R/R_{0}$ is plotted as a function of the radial length. Inset shows the slope of the main plot. The density is divided by $R_{0}$ rather than $R_{0}^{2}$ for clarity. As the compaction decreases, the influence gradually propagates to the interior and causes the mass density distribution to approach homogeneity. The slope with compaction above 0.43 is small because the sample has not entered the crumpling regime and remains as a sheet. The error bar in the inset is drawn for $R/R_{0}$=0.17 from three more samples.} \label{mass}
\end{center}
\end{figure}

An extensive investigation of the mass distribution is done to reveal how the individual mass relaxes locally and globally throughout the assembling. A simple coarse-grained method is first adopted to calculate the mass density distribution. After the radius of gyration, $R_{g}^{global}=r_{max}$, is determined, we measure the density along the radial direction and  average over azimuthal angles. 
This simple method, however, suffers from possible interference from the deviation of the sample from a perfect sphere. To remedy this defect, we employ another more detailed investigation in which the radius of gyration is defined locally as $R_{g}^{loc}=r_{max}^{loc}$ where $r_{max}^{loc}$ is determined by the final vanishing of particles within an interval of solid angle. We divide the whole sample into $20\times 36\times10$ pieces along the radial, azimuthal, and tilting angles, respectively. After all the masses have been accumulated in each piece, it is divided by the volume of the piece to give the corresponding density,  $\rho$. In contrast to a homogenous distribution, this second method detects a linear proportionality for low compaction $R/R_0$ sample, see Fig.\ref{mass}.

In the inset of Fig.\ref{mass}, the slope $m=d\rho/dR_{g}^{loc}$ decreases as $R/R_{0}$ shrinks. In other words, the mass tends to accumulate near the crust initially, but eventually shifts homogeneously to the core. This trend as well as that of the fractal dimension points to the physical picture that elastic and plastic deformations dominate at separate stages during the crumpling. This is contrast to the crumpled elastic wires\cite{Neil2} for which there is no second homogeneous stage. Lin {\it et al.}\cite{Neil1} have performed 3D crumpling experiment and measured the number of layers and vertices. They found that these two formations dominate at different stages. In the beginning, layers are created in abundance by the folding. The distribution of facet size is wide, which results in a nonuniform mass distribution. As the ball size decreases, it becomes difficult to generate new layers. Further compression now only serves to bulk the existing layers and induce the permanent or plastic deformations, such as the ridges and vertices. The small facet size that characterizes this stage contributes to a more uniform mass distribution. 
This porous yet compact packing resembles that of a granular media. This analogy has been put forward by  $\AA$str$\ddot{o}$m {\it et al.}\cite{jam} except that our transition  is gradual and happens much earlier.

\section{Discussions}

The calculations for both the fractal dimension and density exponent involve statistical averaging over the bulk of samples. The fact that each crumpled sheet exhibits thousands upon thousands vertices convinces us that it has gone through numerous schochastic processes. According to the central limit theorem, the random errors they create will have cancelled each other considerably and reached rather accurate properties. Similar practice can be found in Ref.\cite{ET}.

Microscopically, we can imagine the thin sheet as a network of triangular meshes consisting of springs, such as the tethered membrane model by Kantor, Kardar, and Nelson\cite{Nelson_book}. In the early stage, the strain is below the threshold of plastic deformation and so most springs still obey the Hook's stress-strain relation. The correlation length of the material fully extends and exhibits the same property as an elastic sheet. Only when the volume fraction is high will the springs enter the plastic regime and render the correlation length extremely short. This causes the gradual transition into a structure which is porous and yet homogenous in mass distribution, similar to the granular packing\cite{jam}. 

The volume fraction is estimated to be about 0.45 when the compaction $R/R_0$ equals 0.15 which means that more than half of the interior is still filled with air. This volume fraction is much smaller than the fcc packing, 0.74, and the resistance divergence point, 0.75, in elastic sheets\cite{jam}. One question arises: How does this structure derive its incredible resistance from such an inefficient packing compared to the granular system? One reason is that the system is constantly trapped in metastable states\cite{Balankin1,Nagel} due to the non-Markovian nature of the process. Another conceivable answer is from the material science. It is known in mechanical engineering that a beam buckles under an axial load when its length exceeds 50 to 200 times of its thickness\cite{timoshenko}. For a shorter beam, the deformation switches from the bending modulus to the normally much-larger Young's modulus. In other words, each material has an intrinsic minimum size of facet, below which it takes an enormously large stress. We believe this causes the poor packing of the hollow polyhedrons surrounded by these facets inside the crumpled structure. The surprising thing is that, rather than complicating the structure, the plasticity enables the interior to reach a higher fractal dimension and more homogeneous packing, both of which contribute to its  high mechanical resistance.

\section{Conclusions and outlook}
By employing the X-ray microtomography, we find that the mass distribution inside the aluminum ball is inhomogeneous at low volume fraction and with a fractal dimension slightly larger than 2. We call this the loose-packing regime. This is reminiscent of the same inhomogeneity observed in the wire crumpling system\cite{Herrmann, cw, Neil2} which is without the plasticity. As we compress the ball further, it enters the compact-packing regime where the fractal dimension increases to 2.8 and the mass distribution becomes homogeneous. This finding implies that, with the shrinkage of ball size, the sheet runs out of rooms to generate new layers and thus can only start to buckle. When the strain exceeds the yield point, this creates a sudden surge of ridges and vertices\cite{Neil1} which mark out many separate facets whose size gets smaller as their number increases under further compression. The homogeneous mass distribution and strong resistance of this configuration are analogous to those of granular packing as suggested by previous simulation\cite{jam}.

There was a serious debate on whether the scaling law holds for real material. Imaging different materials would be an exciting outlook. The reason we focused on the aluminum foil was due to its stability. Other material, such like HDPE, Mylar, and paper, all exhibit a slow swelling relaxation\cite{relaxation}. This phenomenon ruins the 3-D image reconstruction during the long period of imaging, which takes up more than three hours. Adhesive or some smart way to confine the sample without introducing further noises may be the future direction to overcome this obstacle.

\section{Acknowledgement}

We benefit from fruitful discussions with Itai Cohen and Gordon J. Berman and technical helps from Yen-Wei Lin. Support by the National Science Council in Taiwan under grant 95-2112-M007-046-MY3 is acknowledged.



\begin{thebibliography}{999}

\bibitem{Herrmann}
N. Stoop, F. K. Wittel and H. J. Herrmann,
Phys. Rev. Lett. {\bf 101}, 094101 (2008).

\bibitem{Witten07}
T.A. Witten,
Rev. Mod. Phys. {\bf 79}, 643 (2007).

\bibitem{Mahadevan99}
E. Cerda, S. Chaieb, F. Melo and L. Mahadevan,
Nature {\bf 401}, 46 (1999);
S. Chaieb, F. Melo and J.-C. G$\acute{e}$minard,
Phys. Rev. Lett. {\bf 80}, 2354 (1998).

\bibitem{Kudrolli}
D. L. Blair and A. Kudrolli,
Phys. Rev. Lett. {\bf 94}, 166107 (2005);
C. A. Andresen, A Hansen, and J. Schmittbuhl,
Phys. Rev. E {\bf 76}, 026108 (2007).

\bibitem{Nelson_book}
D. Nelson, S. Weinberg and T. Piran (eds),
{\em Statistical Mechanics of Membranes and Surfaces} (World Scientific, Singapore, 2004); Y. Kantor, M. Kardar, and D. R. Nelson, Phys. Rev. Lett. {\bf 57}, 791 (1986); Y. Kantor and D. R. Nelson, Phys. Rev. Lett. {\bf 58}, 2774 (1987) and Phys. Rev. A {\bf 36}, 4020 (1987)

\bibitem{Gompper}
G. A. Vliegenthart and G. Gompper,
Nature Mater. {\bf 5}, 216 (2006).

\bibitem{Timonen}
T. Tallinen, J. A.  $\AA$str$\ddot{o}$m and J. Timonen,
Nature Mater. {\bf 8}, 25 (2009).

\bibitem{jam}
J. A. $\AA$str$\ddot{o}$m, J. Timonen and M. Karttunen,
Phys. Rev. Lett. {\bf 93}, 244301 (2004).

\bibitem{Balankin1}
A. S. Balankin, I. C. Silva, O. A. Mart$\acute{i}$nez and O. S. Huerta,
Phys. Rev. E {\bf 75}, 051117 (2007).

\bibitem{Balankin2}
A. S. Balankin, R. C. deOca and D. S. Ochoa,
Phys. Rev. E {\bf 76}, 032101 (2007).

\bibitem{Gomes1}
M. A. F. Gomes, V. P. Brito, A. S. O. Coelho and C. C. Donato,
J. Phys. D {\bf 41}, 235408 (2008);
M. A. F. Gomes, T. I. Jyh, T. I. Ren, I. M. Rodrigues and C. B. S. Furtado,
{\it ibid.} {\bf 22}, 1217 (1989).

\bibitem{Gomes3}
M. A. F. Gomes, C. C. Donato, S. L. Campello, R. E. de Souza and R Cassia-Moura,
J. Phys. D {\bf 40}, 3665 (2007).


\bibitem{Neil1}
Y. C. Lin, Y. L. Wang, Y. Liu and T. M. Hong,
Phys. Rev. Lett. {\bf 101}, 125504 (2008).


\bibitem{Timonen2}
T. Tallinen, J. A.  $\AA$str$\ddot{o}$m and J. Timonen,
Phys. Rev. Lett. {\bf 101}, 106101 (2008).


\bibitem{hwu1}
Y. Hwu, W. L. Tsai, A. Groso, G. Margaritondo and J. H. Je,
J. Phys. D {\bf 35}, 105 (2002)

\bibitem{hwu2}
Y. T. Chen, T. N. Lo, Y. S. Chu, J. Yi, C. J. Liu, J. Y. Wang, C. L. Wang, C. W. Chiu, T. E. Hua, Y. Hwu, Q. Shen, G. C. Yin, K. S. Liang, H. M. Lin, J. H. Je and G. Margaritondo,
Nanotechnology {\bf 19}, 395302 (2008).

\bibitem{seg_book}
J. S. Suri, D. L. Wilson and S. Laxminarayan (eds),
{\em Handbook of Biomedical Image Analysis, Volume 2, Segmentation Models Part B} (Springer, 2005).

\bibitem{Cannell}
D. W. Schaefer, J. E. Martin, P. Wiltzius and D. S. Cannell,
Phys. Rev. Lett. {\bf 52}, 2371 (1984);
D. W. Schaefer,
Nature {\bf 243}, 1023 (1989);
R. Buzio, C. Boragno, F. Biscarini, F. B. De Mongeot and U. Valbusa,
Nature Mater. {\bf 2}, 233 (2003).


\bibitem{stanley} S. Miyazima and H. E. Stanley, Phys. Rev. B {\bf 35}, 8898 (1987).


\bibitem{Neil2}
Y. C. Lin, Y. W. Lin and T. M. Hong,
Phys. Rev. E {\bf 78}, 067101 (2008).

\bibitem{cw}
C. C. Donato, M. A. F. Gomes and R. E. de Souza,
Phys. Rev. E {\bf 66}, 015102(R) (2002);
C. C. Donato, M. A. F. Gomes and R. E. de Souza,
Phys. Rev. E {\bf 67}, 026110 (2003).

\bibitem{Gennes}
P. G. de Gennes, {\em Scaling Concepts in Polymers Physics} (Cornell University Press, Ithaca, NY, 1979).

\bibitem{ET}
S. Deboeuf, M. Adda-Bedia and A. Boudaoud,
Euro. Phys. Lett. {\bf 85}, 24002 (2008).

\bibitem{Nagel}
K. Matan, R. B. Williams, T. A. Witten, and S. R. Nagel,
Phys. Rev. Lett. {\bf 88}, 076101 (2002). 

\bibitem{timoshenko}
S. P. Timoshenko and J. Gere, 
{\em Theory of Elastic Stability} (McGraw-Hill, 1961) 2nd ed..

\bibitem{relaxation}
A. S. Balankin, O. S. Huerta, R. C. M. deOca, D. S. Ochoa, J. M. Trinidad, and M. A. Mendoza, Phys. Rev. E {\bf 74}, 061602 (2006).


\end{thebibliography}
\end{document}